\documentclass[a4paper,11pt]{article}
\usepackage[utf8]{inputenc}

\usepackage{amsfonts}
\usepackage{amssymb}
\usepackage{graphicx}
\usepackage{amsmath}
\usepackage{enumerate}
\usepackage{color}
 \usepackage{multirow}
\usepackage{float}

\RequirePackage{mathrsfs} \RequirePackage[sc]{mathpazo}
\RequirePackage{wasysym} \RequirePackage{setspace}\textheight=650pt
\title{Maxwell's equal-area law for Gauss-Bonnet 
Anti-de Sitter black holes }
\author{A. Belhaj$^{1,2}$, M. Chabab$^2$, H. EL Moumni$^2$, K. Masmar$^2$, M.  B. Sedra$^{3,4}$ \\
\\
{\small $^{1}$D\'epartement de Physique, Facult\'e
Polydisciplinaire, Universit\'e Sultan
 Moulay Slimane, B\'eni Mellal,  Morocco. } \\
{\small $^{2}$High Energy Physics and Astrophysics Laboratory, FSSM,
 \small Cadi Ayyad University, Marrakesh, Morocco.
} \\
{\small $^{3}$  D\'{e}partement de Physique, LHESIR, Facult\'{e} des
Sciences, Universit\'{e} Ibn Tofail,
 K\'{e}nitra, Morocco.} \\
 {\small $^{4}$ Universit\'e Mohammed Premier, Ecole Nationale des Sciences Appliqu\'ees, BP : 3, Ajdir, 32003, Al Hoceima.}
 }
\date{\today }

\begin{document}

\maketitle
\begin{abstract}
Interpreting  the cosmological constant $\Lambda$ as a
thermodynamic pressure and its conjugate quantity as a thermodynamic
volume,  we study  the Maxwell's equal area law of higher dimensional
Gauss-Bonnet-AdS black holes in extended space. These black hole
solutions critically behave  like Van der Waals systems.  It has
been realized that below the critical temperature $T_c$ the stable
equilibrium is violated. We show through numerical calculations that the
critical behaviors  for the uncharged black holes only appear  in
$d=5$. For  the  charged case,  we analyse solutions in  $d = 5$
and $ d = 6$ separately and find that, up to some constrains, 
the critical behaviors  only appear in  the spherical topology. Using  the Maxwell's construction, we also find the isobar line for
which the  liquid-gas-like phases coexist.
\end{abstract}

%\noindent \textbf{Keywords}:  $P$-$V$ criticality, topological AdS
%black holes,  Lovelock-Born-Infeld gravity, Ehrenfest
%thermodynamical  equations.
\newpage

\section{Introduction}

Recently, the study of  thermodynamical properties of  the black
holes using techniques explored in statistical physics  and fluids
has  received a special interest \cite{30,a1,a2}. These researches
have brought new understanding of the fundamental physics associated
with the critical behaviors of   several black holes in various
dimensions using either numerical or analytic
methods\cite{30,4,5,50,6,7,KM}. In particular,  Ads black holes in
arbitrary dimensions have been extensively investigated in many
works
\cite{KM,chin1,Dolan1,Dolan2,GKM,our1,our2,gaussB,eher,Spallucci:2013osa,Spallucci:2013jja,Zhao:2014eja,our,D.
G. Boulware}. More precisely, the state equations $P=P(T,v)$ have
been established   by considering the cosmological constant as the
thermodynamic pressure and its conjugate as the thermodynamic
volume.  In fact, this issue has opened a new way to study the
behavior of the RN-AdS black hole systems using the physics of Van
der Waals fluids. Indeed, it has been shown that the corresponding
$P$-$V$ criticality can  be linked to   the liquid-gas systems of
statistical physics. Moreover, it has been realized  that  the
criticality depends on many parameters including the dimension of
the spacetime \cite{chin1,Dolan1,Dolan2,GKM,our1,our2,gaussB,eher}.

More recently,   a special interest has been devoted to study of
Maxwell's equal for some black hole solutions. More precisely, the
state equations of  various   AdS black hole in $P-v$ diagrams  have
been worked out showing   the existence of undesirable negative
pressure. They appear also  thermodynamic unstable regions
associated with the condition $\frac{\partial P}{\partial v}>0$.  In
this way, the corresponding system  can be contracted and expanded
automatically
\cite{Spallucci:2013osa,Spallucci:2013jja,Zhao:2014eja,Zhao:2014fea,Wei:2014qwa,Zhao:2014owa}.
In  van der Waals  systems,  such  problems have been overcame by
using  the  Maxwell equal area approach.

The aim of this work is to contribute to these researches by
studying higher dimensional uncharged and charged
Gauss-Bonnet-Anti-de Sitter black holes in extended phase space.
Considering the cosmological constant $\Lambda$ as a thermodynamic
pressure and its conjugate quantity as a thermodynamic volume,  we
present   the Maxwell's equal area law of higher Gauss-Bonnet-AdS
black holes in extended space. These black hole solutions involve
critical behaviors   like Van der Waals gas. It has been shown  that
below the critical temperature $T_c$ the stable equilibrium is
violated. Based on  the Maxwell's construction, we obtain  the
isobar line for which the liquid-gas phases coexist.

 The paper is organized as follow.  In  section 2, we
give an  overview on  the thermodynamics of higher dimensional
Gauss-Bonnet black holes in AdS geometry.  In section $3$,   we first 
study the equal area law of Gauss-Bonnet-AdS black hole in extended
phase space for uncharged solutions. Next, we extend the analysis  to the charged case in five dimensions. Then,  we present the
results for higher dimensions. The last section is devoted to
conclusion.

%\vspace{7 cm}

\section{Thermodynamics of Gauss-Bonnet  black holes in AdS space}
In this section, we give an overview on thermodynamics of
Gauss-Bonnet  black holes in AdS space.  This matter is based
on many works given by other authors \cite{gaussB,jhep,RGCai2002,Hendi:2014bba}. Indeed, we start by
considering a  $d$-dimensional
  Einstein-Maxwell theory  in the presence of the  Gauss-Bonnet terms and a cosmological constant
   $\Lambda=-\frac{(d-1)(d-2)}{2l^2}$. The corresponding action reads as
\begin{equation}
\mathcal{I}=\frac1{16\pi}\int d^dx \sqrt{-g}[R-2\Lambda+\alpha_{GB}
 (R_{\mu\nu\gamma\delta}R^{\mu\nu\gamma\delta}-4R_{\mu\nu}R^{\mu\nu}+R^2)-4\pi F_{\mu\nu}F^{\mu\nu}],
 \label{action}
\end{equation}
where $\alpha_{GB}$ is the Gauss-Bonnet coefficient with dimension
$[{\rm length}]^2$. In this action,  $F_{\mu\nu}$ is the Maxwell
  field strength  given by  $F_{\mu\nu}=\partial_\mu A_\nu-\partial_\nu A_\mu$
    where  $A_\mu$ is an abelian gauge field.  Roughly speaking,  the
    discussion    will be given   here corresponds to
   the case with a positive Gauss-Bonnet coefficient,
  namely,  $\alpha_{GB}\geq0$.  As shown in  \cite{jhep}, dynamical
  solutions appear only  in  higher dimensional theories ($d \ge 5$). For
  this reason, it should be interesting to concentrate on  such models.
  In this way, the  above  action  produces a  static black hole solution with  the following metric
\begin{equation}
ds^2=-f(r)dt^2+f^{-1}(r)dr^2+r^2h_{ij}dx^idx^j,
 \label{ds1}
\end{equation}
 where $h_{ij}dx^idx^j$ is  the line element of a $(d-2)$-dimensional maximal
  symmetric Einstein space with the  constant curvature $(d-2)(d-3)k$ and volume
  $\Sigma_k$. It is noted that
   $k$ takes three values $1$, $0$ and $-1$, corresponding
    to the spherical, Ricci flat and hyperbolic topology of the black hole horizon,
    respectively. According  to ~\cite{D. G. Boulware,RGCai2002,D. L. Wiltshir,M. Cvetic},
    the metric function $f$ takes the following form
\begin{equation}
f(r)=k+\frac{r^2}{2\alpha}\left (1-\sqrt{1+\frac{64\pi\alpha
M}{(d-2)\Sigma_k r^{d-1}}-\frac{2\alpha Q^2}{(d-2)(d-3)r^{2d-4}}-
\frac{64\pi\alpha P}{(d-1)(d-2)}} \right ),
\label{fr}
\end{equation}
where $\alpha=(d-3)(d-4)\alpha_{GB}$. In this solution,
$M$ represents  the black hole mass,
 $Q$ is linked  to the charge of the black hole and $P=-\frac{\Lambda}{8\pi}$.
For  well-defined vacuum solution associated with  $M=Q=0$, the
effective Gauss-Bonnet coefficient $\alpha$ and pressure
   $P$ must satisfy the following constraint
\begin{equation}
0\leq\frac{64\pi\alpha P}{(d-1)(d-2)}\leq 1.
\label{al}
\end{equation}
 It is recalled that the mass  $M$ can be given  in terms
 of the horizon  radius $r_h$  of the black hole determined by the largest real
 root of the equation $f(r_h)=0$. It takes the following form
\begin{equation}
M=\frac{(d-2)\Sigma_k r_h ^{d-3}}{16\pi}\left (k+\frac{k^2\alpha}
{r_h^2}+\frac{16\pi P r_h^2}{(d-1)(d-2)}\right )+\frac{\Sigma_k Q^2}{8\pi (d-3)r_h^{d-3}}.
\label{M}
\end{equation}
Similarly,  the Hawking temperature of the black hole  reads as
\begin{equation}
T=\frac1{4\pi}f'(r_h)=\frac{16\pi P r_h ^4/(d-2)+(d-3)k r_h ^2+(d-5)k^2 \alpha-\frac{2 Q^2}{(d-2)r_h ^{2d-8}}}{4\pi r_h  (r_h ^2+2k\alpha)}.
\label{T}
\end{equation}
It is worth noting that in  the discussion of  the thermodynamics of
the black hole in the
 extended phase space by  considering the pressure $P=-\frac{\Lambda}{8\pi}$,
  the black  hole mass $M$  should be identified with  the enthalpy $H\equiv M$
   rather than the internal energy of the gravitational system~\cite{Kastor:2010gq}. It
    turns out that  many other thermodynamic quantities can be obtained using  thermodynamical  equations.
For instance,  the entropy $S$, thermodynamic volume $V$ and
electric potential (chemical potential) $\Phi$ take the following
forms
\begin{equation}
S=\int_0^{r_h} T^{-1}(\frac{\partial H}{\partial r})_{Q,P} dr=
\frac{\Sigma_k r^{d-2}_h}{4}\left (1+\frac{2(d-2)\alpha k}{(d-4)r_h^2}\right ),
\label{S}
\end{equation}

\begin{equation}
V=(\frac{\partial H}{\partial P})_{S,Q}=\frac{\Sigma_k r_h ^{d-1}}{d-1},
\label{vol}
\end{equation}
\begin{equation}\label{Qphi}
    \Phi=(\frac{\partial H}{\partial Q})_{S,P}=\frac{\Sigma_k Q}{4\pi (d-3)r_h^{d-3}}.
\end{equation}
These thermodynamic quantities
 satisfy the following differential form
\begin{equation}\label{diff-Small}
dH= TdS +\Phi dQ +VdP +\mathcal{A} d\alpha,
\end{equation}

where
\begin{equation}\label{beta}
    \mathcal{A} \equiv (\frac{\partial H}{\partial\alpha})
    _{S,Q,P}=\frac{(d-2)k^2\Sigma_k}{16\pi} r^{d-5}_h- \frac{(d-2)k\Sigma_k T}{2(d-4)}r_h^{d-4}
\end{equation}
is the conjugate quantity to the Gauss-Bonnet coefficient $ \alpha$,
being  considered  as a variable.
   By the scaling argument,  the generalized Smarr relation for the
black holes can be written as
\begin{equation}\label{Smarr}
    (d-3)H=(d-2)TS-2PV+2\mathcal{A}\alpha+(d-3)Q\Phi.
\end{equation}
Moreover, many   solutions treating other black holes  are given
 in~\cite{GKM,Kastor:2010gq}.    This class of   black hole solutions  has
 volume $V\sim \Sigma_k r^{d-1}_h$ and an
 area given by
$A\sim \Sigma_k r_h^{d-2}$. In this way,  the black hole horizon has
a  scalar curvature
 $R_h \sim k/r_h^2$. The Gauss-Bonnet term on the horizon is  $R_{\rm GB} \sim
 k^2/r^4_h$. In fact,
  the first term in (\ref{beta}) can be put in  the form $VR_{\rm GB}$, while
   the second term  takes the form $ T A R_h$.
  It is observed that both terms vanish in the flat solution
  corresponding to $k=0$.  It is noted that the first term in (\ref{beta}) is nothing but   the second term in
the black hole mass
      (\ref{M}), while the second term in (\ref{beta}) is identified with  the second term of the
       black hole entropy (\ref{S}) multiplied by Hawking temperature $T$.
      Using  Legendre transformations, the Gibbs free energy and the  Helmholtz free
      energy read as
\begin{equation}
G=G(T,P,Q)=H-TS, \hspace{0.5cm} F=F(T,V,Q)=G-PV.
\label{Gibbs1}
\end{equation}
However, it is recalled that the Helmholtz free energy $F$ can be
obtained by removing  the contribution of the background of the AdS
vacuum solution. In this way,  only the situation associated   with
a negative $F$ can be  regarded as the black hole solution
considered as a thermodynamically  AdS vacuum solution. In the case associated 
with  the hyperbolic horizon
 ($k=-1$)~\cite{T. Clunan}, the black hole entropy
     \eqref{S} could be  negative.  In fact, a negative entropy has no meaning  in
  statistical physics. For these reasons, the constaints
\begin{equation}
F  \le 0,\hspace{0.3cm}S\geq0,\hspace{0.3cm}r_h>0,\hspace{0.3cm}T\geq0,\hspace{0.3cm}
0\leq\frac{64\pi\alpha P}{(d-1)(d-2)}\leq 1.
\end{equation}
should be added.  It is noted that  in the case $k=-1$,  the metric
function $f$ (\ref{fr}) shows the existence of
  a minimal horizon radius $r_h^2 \ge 2\alpha$.
However, the non-negative values  of the black hole entropy
 (\ref{S}) produce  a more strong constraint on the horizon radius
given by  $r^2_h \ge (2+ 4/(d-4))\alpha$.

\section{The equal area law of Gauss-Bonnet-AdS black hole in extended phase space.}

Having given the essentials on the thermodynamics of the
Gauss-Bonnet-Anti-de Sitter black holes, we move to investigate the
corresponding Maxwell's equal area law.
\subsection{The construction of equal area law in $P-v$ diagram}

In the study of  der Waals system, for  an isotherm below the
critical temperature $T< T_c$  the two points, of the plan $(P,v)$,
solving the equation
\begin{equation}
\frac{\partial P}{\partial v}=0
\end{equation}
indicate the the  stability limit of   the system.   It is observed
that the critical  point corresponding to the largest volume is
interpreted as the   stability limit  of the gaseous phase. However,
the critical  point  associated with  the smallest volume
corresponds to the  stability limit  of the liquid phase. The
chemical potential of  such   thermodynamic systems should satisfy
\begin{equation}
d\mu=-S dT+ V dP.
\end{equation}
In the isotherm transformation,  the difference of chemical
potential between two states  with  the pressure $P$ and $P_0$ should
have the following form
\begin{equation}\label{2.2}
\mu-\mu_0=\int_{P_0}^P VdP.
\end{equation}
It is realized  that the Gauss-Bonnet-Ads black holes present
similar behaviors  as  illustrated  in  figure 1.
%\vspace{12 cm}

\begin{center}
\begin{figure}[!th]
\begin{center}
{\includegraphics[scale=1.2]{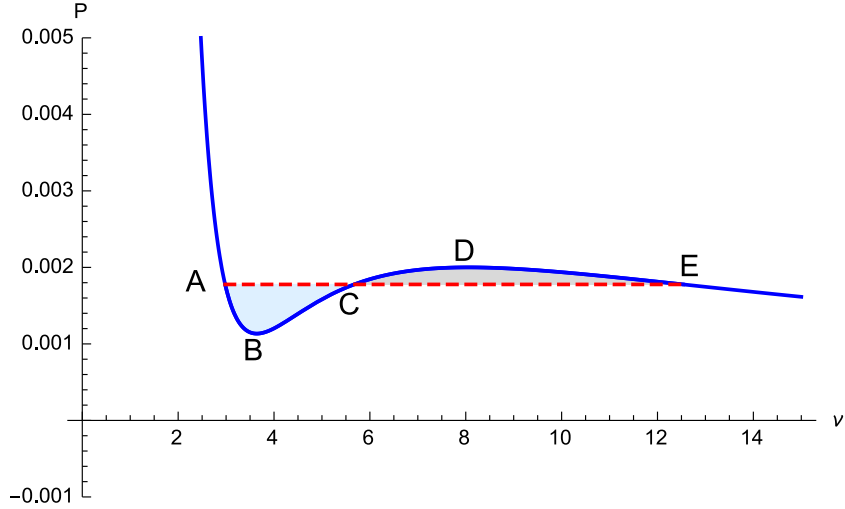}}
\end{center}
 \caption{ A $P-v$ curve below the critical temperature.} \label{fig1} \vspace*{-.2cm}
\end{figure}
\end{center}

It observed  from this figure that,  at the point $E$, the black
hole lies at "gas" phase. However,   at point "A", the black hole
lies at "liquid" phase completely.  Moreover, the region between $A$
and $E$ can be considered as a  coexistence phase. However,  the
oscillating part of the curve between $A$ and $E$ cannot be the
coexistence line due to the  fact that   the part $BD$ violates the
 equilibrium conditions. At  $A$ and $E$, the
chemical potentials $\mu_A (T,P)$ and $ \mu_B (T,P)$  are
 \begin{equation}\mu_A (T,P) =\mu_B (T,P).
 \end{equation}
 This relation produces  the thermodynamic condition for the phase equilibrium. Indeed,
  the equation $(\ref{2.2})$   gives
\begin{equation}
\int_{EDCBA} v dP =0,
\end{equation}
showing  that the area $EDC$ and $ABC$  are equal.\\

In what follows, the main aim is  to  investigate  the position of
the points $A$ and $E$ for Gauss-Bonnet-AdS black holes in higher
dimensions. First, we discuss  the uncharged case then we will
present numerical  solutions  for the  charged black holes in the
next section.

\subsection{Uncharged solutions }
In this subsection, we discus the neutral case corresponding to
$Q=0$. Identifying
  the pressure with the cosmological constant  and the corresponding conjugate
  thermodynamic volume,  the Hawking temperature (\ref{T}) can be
  explored to write down    the state  equation. The latter reads
  as
\begin{equation}
P=\frac{d-2}{4r_h}(1+\frac{2k\alpha}{r_h^2})T-\frac{(d-2)(d-3)k}{16\pi r_h^2}
-\frac{(d-2)(d-5)k^2\alpha}{16\pi r_h^4}.
\label{PQ}
\end{equation}
To make contact with the van der Waals equation, we  should use a
series expansion  with the inverse of specific volume $v$. The
calculation leads to
\begin{equation}\label{Van}
    P=\frac T{v-b}-\frac a{v^2}\approx\frac T v+\frac{bT}{v^2}-\frac a{v^2}+O(v^{-3}).
\end{equation}
Identify the specific volume $v$  with the horizon radius of the
black
 holes as proposed in ~\cite{GKM,KM,our}
\begin{equation}\label{sv}
    v=\frac{4r_h}{d-2},
\end{equation}
 the specific volume $v$  can be related  to the horizon radius
 $r_h$. Using~(\ref{sv}),   one can recover all the results
   in terms of $v$. In this way, the Eq. (\ref{PQ})  becomes
\begin{equation}\label{states}
P=\frac{T}{v}+\frac{k (3-d) }{\pi
    (d-2) v^2}+\frac{32 \alpha  k T}{(d-2)^2 v^3}-\frac{16 \alpha  k^2(d-5) }{\pi  (d-2)^3
    v^4}.
\end{equation}
To obtain  the equal area isobar, $P=P_0$, one  use   the following
relations:
\begin{equation}
 \Delta \mathcal{S}_1=P_0(v_2-v_1)
\end{equation}
and
\begin{equation}
 \Delta \mathcal{S}_2=\int^{v_2}_{v_1} P dv.
\end{equation}
The equal area law requires the following equality
\begin{equation}
 \Delta \mathcal{S}_1=\Delta \mathcal{S}_2
\end{equation}
leading  to
\begin{eqnarray}\nonumber
P_0 (v_2-v_1)&=&T_0 \ln\left(\frac{v_2}{v_1}\right)+ \frac{ k(d-3)
   }{\pi(d-2)}\left(\frac{1}{v_2}-\frac{1}{v_1}\right)
   - \frac{16\alpha k}{\pi (d-2)^2}\left(\frac{1}{v_2^2}-\frac{1}{v_1^2}\right)\\&+&\frac{16 \alpha
     (d-5) k^2 (v_1-v_2) \left(v_1^2+v_1
   v_2+v_2^2\right)}{ 3 \pi  (d-2)^3v_1^3 v_2^3}.
   \end{eqnarray}
In the isothermal curves,  the points $v_1$ and $v_2$  should
satisfy
\begin{equation}
 P_0=\frac{T_0}{v_1}+\frac{k (3-d) }{\pi
    (d-2) v_1^2}+\frac{32 \alpha  k T}{(d-2)^2 v_1^3}-\frac{16 \alpha  k^2(d-5) }{\pi  (d-2)^3 v_1^4}
\end{equation}
\begin{equation}
P_0=\frac{T_0}{v_2}+\frac{k (3-d) }{\pi
    (d-2) v_2^2}+\frac{32 \alpha  k T}{(d-2)^2 v_2^3}-\frac{16 \alpha  k^2(d-5) }{\pi  (d-2)^3
    v_2^4}.
\end{equation}
From these equations, one can derive the relations
\begin{equation}
T_0 v_1^3v_2^3- \frac{(d-3) k v_1^2 v_2^2 (v_1+v_2)}{\pi  (d-2)}+\frac{32 k \alpha   T_0 v_1 v_2 \left(v_1^2+v_1
   v_2+v_2^2\right)}{(d-2)^2}= \frac{16 k^2  \alpha  (d-5) (v_1+v_2)
   \left(v_1^2+v_2^2\right)}{\pi  (d-2)^3}
\end{equation}
and
\begin{equation}
 2P_0=T_0\left(\frac{1}{v_1}+\frac{1}{v_2}\right)+\frac{k (3-d) }{\pi
    (d-2) } \left(\frac{1}{v_1^2}+\frac{1}{v_2^2}\right)+\frac{32 \alpha  k T}{(d-2)^2}\left(\frac{1}
    {v_1^3}+\frac{1}{v_2^3}\right)-\frac{16 \alpha  k^2(d-5) }{\pi  (d-2)^3 }\left(\frac{1}
    {v_1^4}+\frac{1}{v_2^4}\right)
\end{equation}
Putting  $x=\frac{v_1}{v_2}$,($0\leq x\leq 1$), we get the following
relations
\begin{eqnarray}\label{a}
 P_0v_2^4x^{3}(1-x)&=& -T_0 v_2^3 x^3 \ln(x) -\frac{(d-3) k v_2^2 (1-x) x^2}{\pi  (d-2)}\\
 \nonumber&+& \frac{\alpha  k (x-1) \left(16 (d-5) k \left(x^2+x+1\right)-48 \pi  (d-2) T_0
   v_2 x (x+1)\right)}{3 \pi  (d-2)^3}
 \end{eqnarray}
and
\begin{equation}\label{b}
0=T_0v_2^3 x^3-\frac{(d-3) k v_2^2 x^2 (x+1)}{\pi  (d-2)}-\frac{16 \alpha
  (d-5) k^2 (x+1) \left(x^2+1\right)}{\pi  (d-2)^3}+\frac{32 \alpha  k T_0 v_2 x \left(x^2+x+1\right)}{(d-2)^2}
\end{equation}
and
\begin{eqnarray}\label{c} \nonumber
 2 P_0 v_2^2 x^2 &= &T_0v_2^3 x^3(1+x)-\frac{(d-3) k v_2^2 x^2
  \left(x^2+1\right)}{d-2}+\frac{32 \pi  k \alpha  T_0 v_2 \left(x^4+x\right)}{(d-2)^2}\\
&-& \frac{16  k^2 \alpha  (d-5)  \left(x^4+1\right)}{(d-2)^3}.
 \end{eqnarray}
Using  Eqs (\ref{a}),(\ref{b}) and(\ref{c}),  we get  a polynomial
equation admitting  $v_2^2$  as a  real positive root
\begin{eqnarray}\label{d}
a\; v_2^4+b\; v_2 ^2+c=0
\end{eqnarray}
where the coefficients $a$, $b$ and  $c$ read as
\begin{eqnarray}
a&=&3 (d-3) (d-2)^4 k x^5 (-2 x+(x+1) \ln (x)+2),\\\nonumber
b&=& 16 \alpha  (d-2)^2 k^2 x^3 \left(3 (d-5) (x+1) \left(x^2+1\right) \ln (x)-(x-1) \left((7
   d-29) x^2-2 (d+1) x+7 d-29\right)\right) \\ \nonumber
c&=&256 \alpha ^2 (d-5) k^3 (x-1) \left(-3 \alpha +x \left(3 x^5+4 x^4-3 \alpha  \left(x
   (x+1) \left(x^3+x+1\right)+1\right)+5 x^3+5 x+4\right)+3\right).
\end{eqnarray}
Solving the above polynomial  equation,  we get
\begin{equation}\label{v22}
v_2^2=\frac{-b+\sqrt{b^2-4a c}}{2a}.
\end{equation}
In the limit $x\rightarrow1$, one should have $v_1=v_2=v_c$. In what
follows, we   discuss the convergence of this limit in terms of the
topology  and  the   dimension of the space-time.  In fact,   three
situations can appear. They are classified as follows:
\begin{itemize}
\item In the case of ${\bf k=0}$ corresponding to  flat topology,   this limit diverges for  any dimensions $d$. This shows the absence of the critical points. Indeed, this can
be seen  from the equation of state eq. (\ref{states}).   The letter  reduces to $P=\frac{T}{v}$ indicating   that  no phase
transition can happen.
\item In the case of ${\bf k=-1}$ associated with  the  hyperbolic topology,  a
negative value of  $v_2^2$  appears in  $d=5$. For $d\geq 6$, the
limit also diverges. This implies that  there does not exist any
phase transition.
\item In the case of ${\bf k=1}$ corresponding to  the spherical topology,  this limit diverges  only when $d\geq 6$. In the case of $d=5$, we have
\begin{equation}\label{225}
v_c=\lim_{x\rightarrow 1}\sqrt{\frac{y_1}{y_2}}= 4 \sqrt{\frac{2}{3}\alpha},\;with\quad
 \left\{\begin{array}{r c l}y_1&= & 32 \alpha  (x-1)^3\\y_2&= &18 x^2 (x+1) \ln (x)-36
 (x-1) x^2.\end{array}\right.
 \end{equation}
\end{itemize}
It is noted that a similar result has been found in  \cite{jhep}.
Substituting   the ($\ref{225}$) in ($\ref{b}$) to eliminate $v_2$
and setting $T_0=\chi T_c$,  the critical temperature reads as
\begin{equation}
T_c=\frac{1}{2\pi \sqrt{6\alpha}}.
\end{equation}
We can also obtain  the following relation
\begin{eqnarray}\nonumber
512 \sqrt{\frac{2}{3\alpha}} \chi x^3 \left(\frac{y_1}{y_2}\right)
(x^3-1)+\frac{2048}{9}\left( 8 \sqrt{\frac{2\alpha}{3}}(x^3-1)+
\left(\frac{y_1}{y_2}\right)^{\frac{1}{2}} (1-x^2).
\right)=0.
\end{eqnarray}
In the table $(\ref{tab1})$,  we   present  the numerical values of
the $v_{1,2}$ and $P_0$ for different values   of the constant $\alpha$.

% \usepackage{multirow}
\begin{centering}
\begin{table}[H]
\begin{centering}
\begin{tabular}{|l|l|l|l|l|l|}
\hline
\textit{\textbf{$\alpha$}} & \textit{\textbf{$\chi$}} & \textit{\textbf{$x$}} & \textit{\textbf{$v_1$}} & \textit{\textbf{$v_2$}} &  \textit{\textbf{$P_0$}}\\ \hline\hline
\multirow{3}{*}{$\alpha=0.5$}
& 1 & 1 & 2.3094  & 2.3094 &0.01326\\ \cline{2-6}
 &0.8 &0.07111  & 0.85537  & 12.0282 &0.00471\\ \cline{2-6}
 &0.7& 0.02543 &  0.67313 &26.4629  &0.00213\\ \hline\hline
 \multirow{3}{*}{$\alpha=1$}
&  1 & 1 & 3.2659  &3.26599 &0.00663\\ \cline{2-6}
& 0.8 & 0.07111 &  1.20969 & 17.0104&0.00235 \\ \cline{2-6}
 &0.7& 0.02543 & 0.95195  &37.4241 & 0.00106\\ \hline\hline

 \multirow{3}{*}{$\alpha=2$}
 &1 & 1 & 4.6188  & 4.6188 &0.00331 \\ \cline{2-6}
& 0.8  & 0.07111 & 1.71076  & 24.0564 &0.00117 \\ \cline{2-6}
& 0.7 & 0.02543 & 1.34627  & 52.9257 & 0.00053\\ \hline\hline

\end{tabular}
\caption{Numerical solutions for $x$, $v_1$, $v_2$ and $P_0$  at different
 temperature with  $Q=0$  in  five dimension with the  spherical topology.}
\label{tab1}
\end{centering}
\end{table}
\end{centering}

From the table (\ref{tab1}),  we can see that $x$ is  not linked  to
the coupling constant $\alpha$. However, it  increases  with  at
certain values of  $\chi$.  The specific volume $v_2$ decreases and
increases  with $\chi$ and
 $\alpha$ respectively. The pressure $P_0$ increase   and decrease  with $\chi$ and
 $\alpha$ respectively. For more details,  we plot the pressure $P$ at constant temperature
in terms of the specific
 volume $v$ for  different values of the coupling constant $\alpha$.

\begin{center}
\begin{figure}[H]
  \begin{tabbing}
\hspace{9cm}\=\kill
\includegraphics[scale=.9]{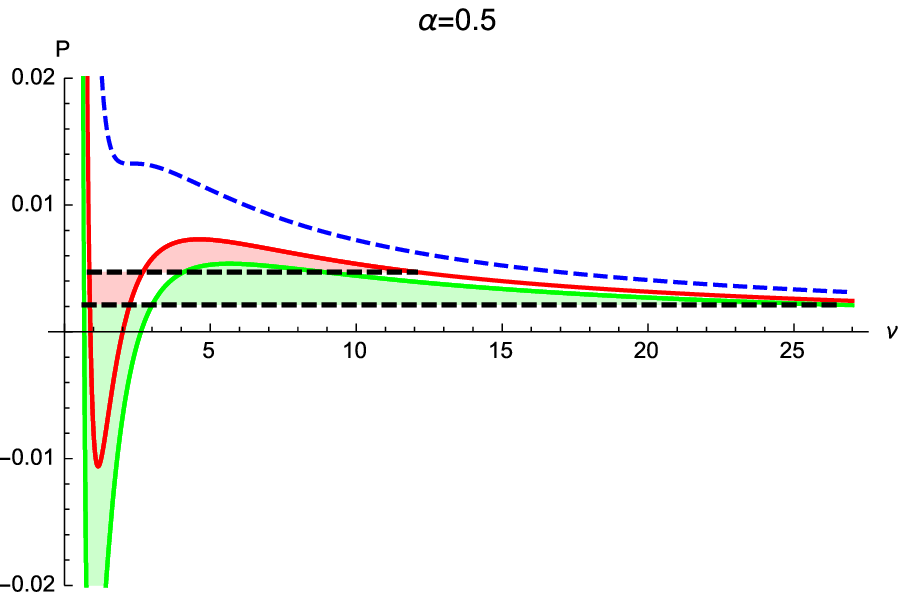} \>\includegraphics[scale=.9]{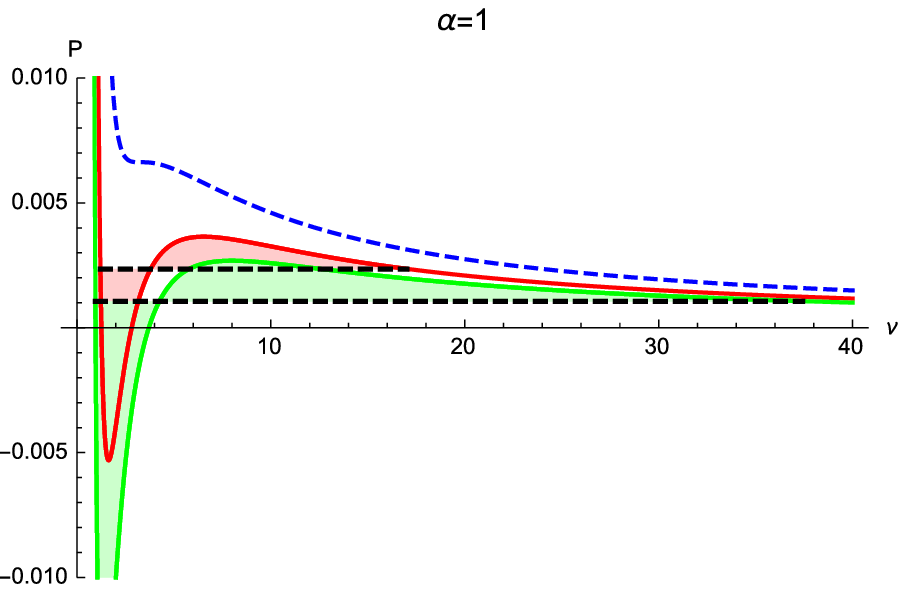} \\
\includegraphics[scale=.9]{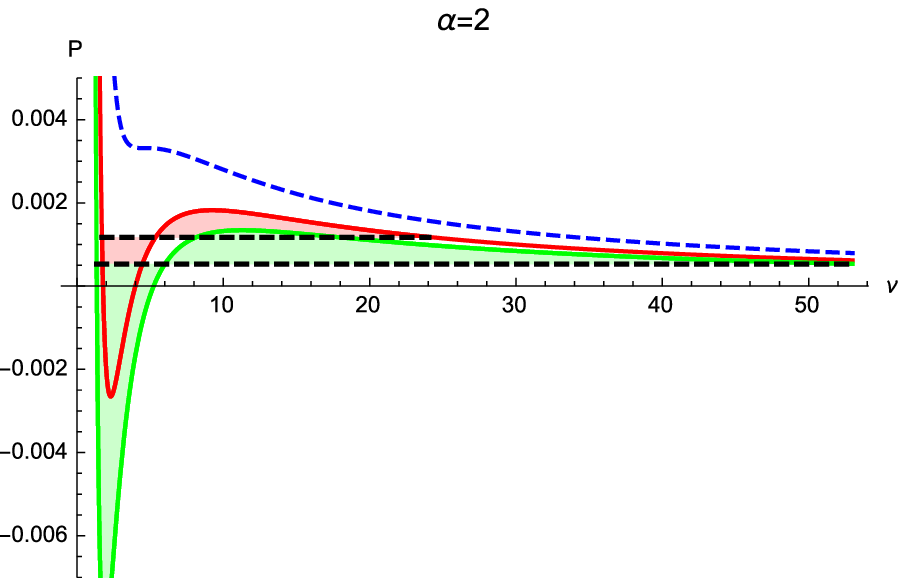}%\>\includegraphics[scale=.9]{11.eps}
\end{tabbing}
\caption{The $P-v$ diagram uncharged  Gauss-Bonnet-AdS black holes in five dimension.  The dashed blue
curve corresponds  to the critical temperature $T_c$, the red one is associated with the  isotherm with $ 0.8 T_c$
 and the green one corresponds  to $0.7 T_c$.  } \label{ffig2} \vspace*{-.2cm}
\end{figure}
\end{center}

For different values of $\alpha$, we observe that the isobar in the
isotherm will be short  by increasing the  temperature.  When the
temperature reaches the critical one,  the boundaries of the isobar
coincide $v_1=v_2=v_c$.

\section{ Charged black hole solutions}
This section concerns  the case of    charged Gauss-Bonnet-AdS black
holes in the  AdS geometry. In this case,  the state equation  can
be written as
\begin{equation}\label{xxxz}
P=\frac{T}{v}+\frac{k (3-d) }{\pi
    (d-2) v^2}+\frac{32 \alpha  k T}{(d-2)^2 v^3}-\frac{16
     \alpha  k^2(d-5) }{\pi  (d-2)^3 v^4}+ \frac{2^{4 d-11}  (d-2)^{4-2 d} Q^2}{\pi
     v^{2d-4}}.
\end{equation}
A close inspection on this equation shows that  one may give a
separate study. First,  we deal with the five dimensional case. Then,
we study   models associated with   $ d \geq 6$.
\subsection{ Five dimensional case}
In  $d=5$,  the equation of the state (\ref{xxxz}) reduces to
\begin{equation}
P=\frac{T}{v} \left(1+\frac{32 \alpha  k}{9 v^2}\right)-\frac{2 k}{3 \pi  v^2}+\frac{512 Q^2}{729 \pi
   v^6}.
\end{equation}
Using a similar analysis used in the case of  uncharged solutions,
the equations (\ref{a}), (\ref{b}) and (\ref{c}) become respectively
\begin{equation}
P_0 v_2^6 (1-x) x^5= -\frac{2 k
   v_2^4 x^4}{3 \pi }+\frac{512 Q^2}{3645 \pi }+ \left(\frac{2 k v_2^4}{3 \pi }-\frac{512 Q^2}{3645 \pi }\right) x^5 +T_0 \left(-\frac{16}{9} \alpha  k v_2^3 x^5+\frac{16}{9} \alpha  k v_2^3
   x^3-v_2^5 x^5 \ln (x)\right)
\end{equation}
\begin{eqnarray}\nonumber
&& T_0 \left(96 \pi  \alpha  (d-2)^3 k v_2^3 \left(1+x+x^2\right) x^3+3 \pi
   (d-2)^5 v_2^5  x^5\right)-6 (d-2)^4 k v_2^4  (x+1) x^4\\ &+&512 Q^2
   \left(1+x)\left(1+x^2+x^4\right)\right)=0,
\end{eqnarray}
and
\begin{equation}
2 P_0 v_2^6 x^6=T_0 \left(\frac{32}{9} \alpha  k v_2^3 \left(x^3+1\right) x^3+v_2^5
   (x+1) x^5\right)+\frac{512 Q^2 \left(x^6+1\right)-486 k v_2^4 x^4 \left(x^2+1\right)}{729 \pi }.
\end{equation}

Using  the above equations, we obtain also  the  polynomial equation
of  $v_2$. It is given by
\begin{equation} \label{poly}
a v_2^6+ bv_2^4 + c v_2^2+ \gamma =0
\end{equation}
where the coefficients $a$, $b$ and $c$ take now the following form
\begin{eqnarray}
a&=& 7290 k (x-1) x^6+3645 k (x+1) x^6 \ln \left(\frac{1}{x}\right),\\
b&=& 6480 \alpha  k^2 (x-1)^3 x^4,\\
c&=&-4608 Q^2 \left(x^5-1\right) x^2-3840 Q^2 (x+1) \left(x^4+x^2+1\right) x^2 \ln
   \left(\frac{1}{x}\right),\\
\gamma&=&4096 \alpha  k Q^2 (x-1)^3 \left(x^2+x+1\right) (x (x+3)+1).
\end{eqnarray}
The limit $x\rightarrow1$, associated  with  $v_1=v_2=v_c$,  leads
to
\begin{equation}\label{x6}
81 k v_c^6+864 \alpha  k^2 v_c^4-+1280 Q^2 v_c^2+8192 \alpha  k
Q^2=0.
\end{equation}
This shares  similarities  with the one obtained in  \cite{jhep}.
The critical specific volume,  which is the positive
 real root of (\ref{x6}),  takes the following form

\begin{equation}\label{vc}
v_c^2=\frac{32 \alpha }{9}+\frac{80 Q^2}{9 \sqrt[3]{3} X}+\frac{64 k^3\alpha ^2}{3 \sqrt[3]{3}
   X}+\frac{16 X}{9 k\ 3^{2/3}}
\end{equation}
with
\begin{equation}
X=\sqrt[3]{72 \alpha ^3 k^6+126 \alpha  k^3 Q^2+\sqrt{3} \sqrt{3888 \alpha ^4 k^9 Q^2+4392
   \alpha ^2 k^6 Q^4-125 k^3 Q^6}}.
\end{equation}
 The  corresponding critical temperature reads as
 \begin{equation}
 T_c=\frac{4 \left(81 k v_c^4-256 Q^2\right)}{81 \pi  v_c^3 \left(32 \alpha  k+3
   v_c^2\right)}.
 \end{equation}
As in the uncharged case,  we  discuss the existence of the critical
points for different topologies. In fact,  we have three situations:
 \begin{itemize}
 \item for the flat topology, the apparition of  $k$,  in fourth term in (\ref{vc}),
 reveals
   the absence of the critical behavior.
 \item for the hyperbolic one,  the constraint on the positivity of
 the temperature and the  specific volume does not show critical behavior.
 \item  for the spherical topology, the existence of  the critical point is controlled by the following constraint
 \begin{equation}
 |Q|\leq 6 \alpha.
 \end{equation}
 \end{itemize}
The  Maxwell's construction can be obtained using the  same analysis
of the previous section. This  gives rise
   to an equation depend only on $x$. To derive    the values of $v_{1,2}$ and the $P_0$, we should find
   $x$. For the charged case, we list  all these results on the  table (\ref{tab2}). 
    \begin{centering}
\begin{table}[H]
\begin{centering}
\begin{tabular}{|l|l|l|l|l|l|}
\hline
\textit{\textbf{$(\alpha,Q)$}} & \textit{\textbf{$\chi$}} & \textit{\textbf{$x$}} & \textit{\textbf{$v_1$}} & \textit{\textbf{$v_2$}} &  \textit{\textbf{$P_0$}}\\ \hline\hline
\multirow{3}{*}{$(1,1)$}
&  1 & 1 & 3.5447  & 3.5447 & 0.00619\\ \cline{2-6}
& 0.8 &0.11658  & 1.69635  & 14.5506 &0.00254  \\ \cline{2-6}
 &0.7&0.05765  & 1.50503  & 26.1024 &0.00139  \\ \hline \hline
 \multirow{3}{*}{$(1,\frac{3}{2})$}
& 1 & 1 & 3.77814  & 3.77814 &0.00582 \\ \cline{2-6}
&0.8 & 0.13991 & 1.95332  &13.961  &0.002531  \\ \cline{2-6}
 &0.7& 0.07388 & 1.759006  & 23.8078 & 0.00146 \\ \hline\hline

 \multirow{3}{*}{$(2,3)$}
 & 1 & 1 & 5.3431  & 5.3434 &0.00291 \\ \cline{2-6}
 &0.8 & 0.13991 &  2.76241 & 19.7438 & 0.00126 \\ \cline{2-6}
 &0.7& 0.07388 &  2.48768 & 33.6694 & 0.00073 \\ \hline\hline

\end{tabular}
\caption{Numerical values of $x$, $v_1$, $v_2$ and $P_0$ at different
 temperature  in five dimension with the  spherical topology  in the presece of the  charge.}
\label{tab2}
\end{centering}
\end{table}
\end{centering}

It follows from  table (\ref{tab2}) that   $P_0$ decreases when one
increases   $\alpha$, $Q$ and $\chi$. However,
the specific volume $v_2 $ increases when $\alpha$  and $\chi$ decrease. In fact,
 the charge decreases the values of  $v_2$. An important remark that
 emerges from this calculation is that  $x$ remains constant where we have a  proportionality between incremented
values of the charge and coupling constant
$(\alpha_1,Q_1)\propto(\alpha_2,Q_2)$ showing that the  charge and
$\alpha$ can be interrelated \cite{jhep}.
 In what follows,  we plot the isotherms in the $(P,v)$ diagram and show the equal Maxwell's
 area.
 \begin{center}
\begin{figure}[H]
  \begin{tabbing}
\hspace{9cm}\=\kill
\includegraphics[scale=.9]{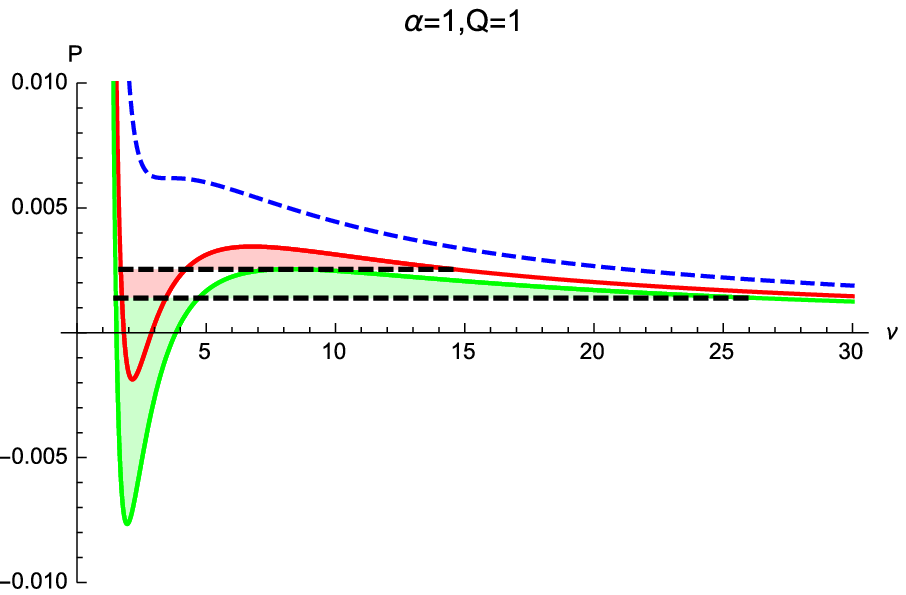} \>\includegraphics[scale=.9]{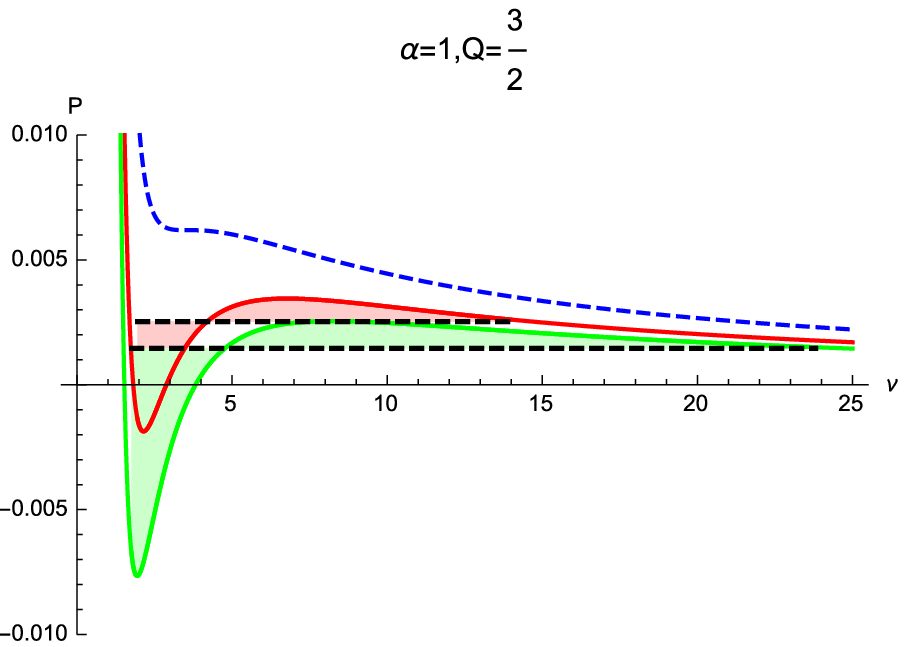} \\
\includegraphics[scale=.9]{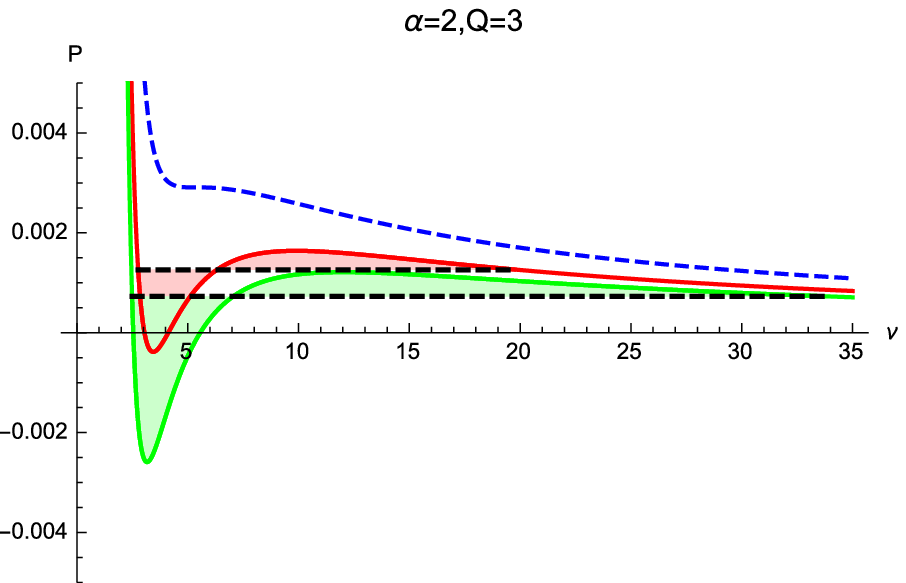}%\>\includegraphics[scale=.9]{11.eps}
\end{tabbing}
\caption{The $P-v$ diagram charged  Gauss-Bonnet-AdS black holes in five dimensions.  The dashed blue
curve corresponds to the critical temperature $T_c$, the red one corresponds to isotherm with $ 0.8 T_c$
 and the green one corresponds  to $0.7 T_c$.  } \label{ffig3} \vspace*{-.2cm}
\end{figure}
\end{center}

%\begin{equation}
%
% \end{equation}

 \subsection{Higher dimensional cases}
 The complete  form  of the equation of  states (\ref{states})  gives also a polynomial
  equation of $v_2$.  Similar calculations can be done for higher
  dimensional black holes. In the limit $x\rightarrow1$, the specific volume is  the real positive solution with  the
 following polynomial form
 \begin{eqnarray}\nonumber
&& \frac{1}{524288}(d-2)^8 Q^2 v_2^8 \left(96 \alpha  (2 d-7) k+(d-2)^2 (2 d-5)
   v_2^2\right) =\\ \nonumber
   & &4^{-2 d} k ((d-2) v_2)^{2 d} \left(12 \alpha ^2
   (d-5) k^2-\frac{3}{4} \alpha  (d-2)^2 k v_2^2+\frac{1}{256} (d-3) (d-2)^4
   v_2^4\right)
 \end{eqnarray}
 The critical temperature  reads  as
 \begin{equation}
 T_c=\frac{2^{4 d-11} Q^{\frac{1}{d-3}} \left((d-2) v_2 Q^{\frac{1}{3-d}}\right)^{7-2 d}
   \left(\frac{16^{3-d} Q^{\frac{6}{d-3}} \left((d-2) v_2
   Q^{\frac{1}{3-d}}\right)^{2 d} \left(32 \alpha  (d-5)+(d-3) (d-2)^2 k
   v_2^2\right)}{(d-2)^8 v_2^8}-2\right)}{\pi  \left((d-2)^2 v_2^2+96
   \alpha  k\right)}
 \end{equation}

\begin{center}
\begin{figure}[H]
  \begin{tabbing}
\hspace{9cm}\=\kill
\includegraphics[scale=.9]{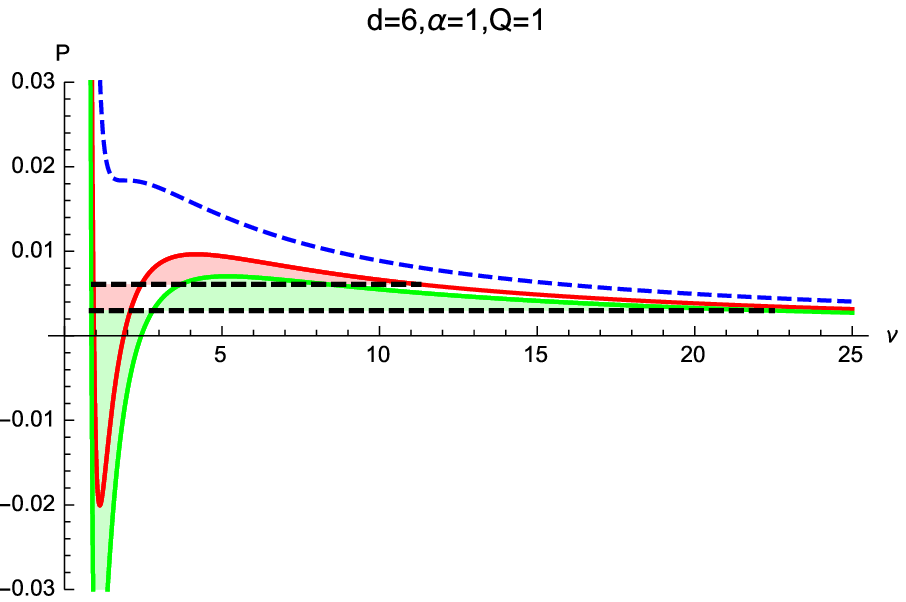} \>\includegraphics[scale=.9]{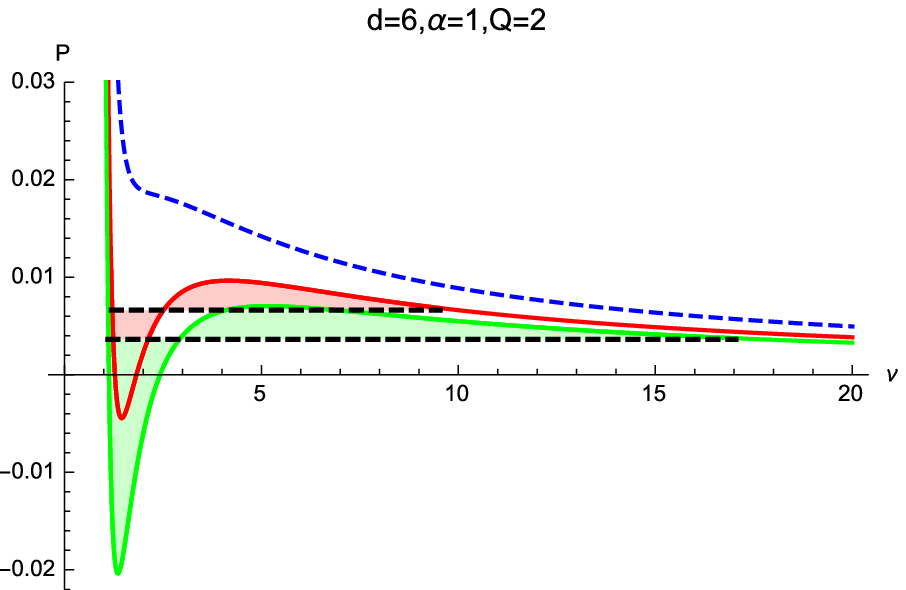} \\
\end{tabbing}
\caption{The $P-v$ diagram of the charged  Gauss-Bonnet-AdS black holes in six dimensions, the dashed blue
curve corresponds to the critical temperature $T_c$, the red one corresponds to isotherm with $ 0.8 T_c$
 and the green one is asscociated with  $0.7 t_c$.  } \label{ffig4} \vspace*{-.2cm}
\end{figure}
\end{center}

It is observed that the Ricci flat topology and the hyperbolic one
do not  allow the existence of critical behaviors.  Up to some
conditions, the critical behaviors appear only in  the case of the
spherical topology corresponding  to $k=1$. In fact, $\alpha
|Q|^\frac{-2}{(d-3)}$ should not be too large \cite{jhep}. For
simplicity reasons, we restrict our study to $d=6$. In particular,
we present the  numerical results in $d=6$ for $x, v_{1,2}$
and $P_0$ in the following table,

  \begin{centering}
\begin{table}[H]
\begin{centering}
\begin{tabular}{|l|l|l|l|l|l|}
\hline
\textit{\textbf{$(\alpha,Q)$}} & \textit{\textbf{$\chi$}} & \textit{\textbf{$x$}} & \textit{\textbf{$v_1$}} & \textit{\textbf{$v_2$}} &  \textit{\textbf{$P_0$}}\\ \hline\hline
\multirow{3}{*}{$(1,1)$}
&  1 & 1 & 1.91384 & 1.91384& 0.0183735\\ \cline{2-6}
& 0.8 &0.08212  & 0.92344  & 11.2444 &0.006092  \\ \cline{2-6}
 &0.7&0.03799  & 0.85358  & 22.4677 &0.002983\\ \hline \hline
 \multirow{3}{*}{$(1,2)$}
& 1 & 1 & 2.17689  & 2.17689&0.002823 \\ \cline{2-6}
&0.8 & 0.12462& 1.18883  &9.53966  &0.0066212  \\ \cline{2-6}
 &0.7& 0.06473 & 1.10356  & 17.0479 & 0.0036418 \\ \hline\hline

\end{tabular}
\caption{Numerical values of $x$, $v_1$, $v_2$ and $P_0$ at different
 values of $\alpha$  in six dimensional spherical topology  with non vanishing charge.}
\label{tab3}
\end{centering}
\end{table}
\end{centering}
It is observed from the table (\ref{tab3}) that
 $P_0$ increases with $Q$ and $\chi$. However, the specific volume $v_2$ decreases with $Q$ and $\chi$.
The  comparison between table (\ref{tab2}) and (\ref{tab3}) shows
the effect of the black hole dimensions. In fact, the $P_0$  and $v_2$  increases  and   decreases  resepectively with the  dimension. To
illustrate this effect,  we plot these results in the
fig.(\ref{ffig4})  showing the Maxwell equal area.

\section{Conclusion}

In this paper,  we have studied  the  the Maxwell's equal area law
of  higher dimensional  Gauss-BonnetAnti-de-Sitter black holes. The
corresponding  critical behaviors share similarity with  van der
Waals one.  We  have shown that this construction can be used  to
eliminate the region of the  violated stable equilibrium
$\frac{\partial P}{\partial v} > 0$. In particular, we have found
  the isobar line in which the two
 real phases coexist.  It has  been realized that this construction
 can be viewed as  a
 simple way to drive the coordinates of the critical points.  We have presented numerical calculations showing   that  the
critical behaviors  for the uncharged black holes appear only when
$d = 5$. For the  charged case,  we have studied  solutions in   $d =
5$ and $ d = 6$ separately and showed that, up to some constrains, the critical behaviors appear
only in  the spherical topology.

\end{document}